\documentclass[
reprint,
twocolumn,
pra,
aps,
superscriptaddress,
nofootinbib
]{revtex4-2}

\usepackage{amsmath,amssymb,amsfonts,bm}
\usepackage{physics}
\usepackage{hyperref}
\usepackage{graphicx}
\usepackage[caption=false]{subfig}
\DeclareMathOperator{\Cov}{Cov}
\DeclareMathOperator{\Var}{Var}
\DeclareMathOperator{\CoE}{CoE}

\begin{document}

\title{Entanglement response to Temperature in Interacting Two-Qubit Thermal States}

\author{Zain H. Saleem}
\affiliation{Mathematics and Computer Science Division, Argonne National Laboratory, Lemont, Illinois 60439, USA}

\author{Iram Saleem}
\affiliation{Independent Researcher, USA}

\begin{abstract}
We investigate the response of entanglement to temperature variations in interacting two-qubit thermal states. For a general two-qubit interaction Hamiltonian, we derive exact expressions for the thermal concurrence, its first and second derivatives with respect to inverse temperature, and the thermal quantum Fisher information. We show that the rate of change of thermal entanglement is bounded by the thermal quantum Fisher information. We further derive a bound relating entanglement curvature and thermal quantum Fisher information, and show that temperature uncertainty induces a loss of entanglement bounded by the same quantity that determines thermometric sensitivity. These results establish thermal quantum Fisher information as a fundamental constraint on the response and robustness of entanglement in interacting two-qubit thermal states.
\end{abstract}

\maketitle

\section{Introduction}

Quantum Fisher information (QFI) plays a central role in quantum metrology by determining the ultimate precision with which physical parameters may be estimated from quantum states \cite{BraunsteinCaves1994,Paris2009,TothApellaniz2014,Pezze2018}. While early developments focused primarily on coherent parameter encoding through unitary dynamics, increasing attention has recently been devoted to quantum metrology in thermal and equilibrium systems. In this setting, the parameter of interest is often the temperature or inverse temperature of a Gibbs state, and the attainable precision is governed by equilibrium fluctuations rather than coherent phase accumulation \cite{Correa2015OptimalThermometry,DePasquale2016ThermalSusceptibility,Abiuso2024EquilibriumMetrology,Mehboudi2019}. Thermal quantum Fisher information has emerged as a fundamental quantity in quantum thermometry, quantum criticality, and the geometric characterization of equilibrium many-body systems \cite{Zanardi2007,Gu2010}.

Entanglement is widely recognized as an important quantum resource for enhanced metrological performance \cite{Pezze2018,Hyllus2012}. For thermal systems, however, the relationship between entanglement and metrological sensitivity is less transparent. Thermal entanglement arises from the redistribution of Gibbs populations among interacting energy levels and has been extensively studied in spin chains, Heisenberg models, and coupled qubit systems \cite{Arnesen2001,Wang2001,Rigolin2004,Sun2003}. In particular, two-qubit thermal states provide a minimal setting in which both entanglement and metrological sensitivity can be characterized exactly through closed analytical expressions. Their simplicity has made them a paradigmatic platform for investigating finite-temperature quantum correlations and their interplay with thermodynamic fluctuations.

Recent work has revealed a close connection between entanglement and quantum metrology through information-geometric constraints on the response of entanglement to parameter variations. For coherently evolving two-qubit pure states, it was shown that the rate of change of entanglement is bounded by the quantum Fisher information \cite{saleem2026quantumfisherinformationspeed}, while the curvature of entanglement and the robustness of entanglement generation against parameter uncertainty are likewise constrained by the same geometric quantity \cite{saleem2026robustness,Saleem2025CoE}. Together, these results suggest that the quantum Fisher information provides a fundamental limit on how rapidly entanglement can change and how sensitive it is to fluctuations in an encoded parameter. An important open question is whether similar connections persist in thermal equilibrium, where parameter dependence arises through Gibbs-state populations rather than coherent dynamical phases.

In this work we address this question for the most general interacting two-qubit system. Since the Gibbs state depends on the interaction strength $g$ and inverse temperature $\beta$ only through the dimensionless combination $\beta g$, we formulate the problem in terms of inverse-temperature estimation, which provides the natural framework for thermal quantum metrology. We derive exact analytical expressions for the thermal quantum Fisher information, concurrence, and the first and second derivatives of concurrence with respect to inverse temperature. We show that the thermal quantum Fisher information is determined entirely by equilibrium energy fluctuations, while the response of thermal entanglement is governed by changes in the occupations of Bell-state populations.

Using these results, we establish information-geometric constraints on both the first and second derivatives of thermal entanglement. We show that the rate of change of thermal entanglement with inverse temperature is bounded by the thermal quantum Fisher information, extending the connection between entanglement response and metrological sensitivity to thermal equilibrium. We further derive a bound relating the curvature of thermal entanglement to the thermal quantum Fisher information and show that the curvature admits a fluctuation-response representation in terms of a connected correlator between Bell-state occupation and energy-fluctuation power. Finally, we demonstrate that uncertainty in the inverse temperature induces a loss of thermal entanglement bounded by the same thermal quantum Fisher information that governs thermometric sensitivity. Together, these results identify equilibrium fluctuations as the common resource underlying the speed and curvature of thermal entanglement, thermometric sensitivity, and robustness to temperature uncertainty in interacting two-qubit systems.

The remainder of this paper is organized as follows. In Sec. II we introduce the two-qubit thermal model and derive the Bell-basis representation of the Gibbs state. Section III reviews thermal entanglement and the concurrence threshold. In Sec. IV we derive the thermal quantum Fisher information and show that it is determined entirely by equilibrium energy fluctuations. Section V analyzes the symmetric logarithmic derivative and the structure of optimal measurements for thermal parameter estimation. In Sec. VI we introduce the thermal speed of entanglement and establish a bound relating its magnitude to the thermal quantum Fisher information. Section VII derives the thermal curvature of entanglement, establishes a corresponding information-geometric bound, and develops its fluctuation-response interpretation in terms of connected thermal correlators. In Sec. VIII we examine the robustness of thermal entanglement under inverse-temperature uncertainty and connect the resulting entanglement degradation to thermal quantum Fisher information. Finally, Sec. IX contains a discussion of the results and their implications for thermal entanglement and quantum metrology.

\section{Model and Thermal State}

We consider the minimal nontrivial setting of two interacting qubits in thermal equilibrium. A generic bilinear interaction may be written as
\begin{equation}
H=\sum_{i,j=x,y,z}J_{ij}\,\sigma_i\otimes\sigma_j ,
\end{equation}
where $J_{ij}$ is a real coupling matrix. Such interactions arise naturally in exchange-coupled spin systems, superconducting qubits, trapped ions, and effective low-energy models of interacting quantum devices.

As discussed in Ref.~\cite{Saleem2025CoE}, any two-qubit interaction of this form can be reduced by local single-qubit rotations to a canonical anisotropic form. Specifically, the coupling tensor admits the singular-value decomposition
\begin{equation}
J = O_A\, \mathrm{diag}(\eta_x,\eta_y,\eta_z)\, O_B^{T},
\end{equation}
with $O_A,O_B\in SO(3)$. Because concurrence and quantum Fisher information are invariant under parameter-independent local unitaries, these rotations preserve both the entanglement structure and the attainable metrological precision. The transformed Hamiltonian therefore provides a canonical representative of the general two-qubit interaction class.

After these local rotations, the Hamiltonian takes the form
\begin{equation}
H(g)=g\widetilde H,
\qquad
\widetilde H=
\eta_x\,\sigma_x\!\otimes\!\sigma_x
+\eta_y\,\sigma_y\!\otimes\!\sigma_y
+\eta_z\,\sigma_z\!\otimes\!\sigma_z ,
\label{eq:H}
\end{equation}
where $\eta_x\ge\eta_y\ge\eta_z>0$ characterize the interaction anisotropy.

Because thermal states depend on the interaction strength and inverse temperature only through the dimensionless combination $\beta g$, estimation of the coupling strength and inverse-temperature estimation are mathematically equivalent. In the present work we set $g=1$ and formulate the problem in terms of inverse-temperature estimation, which provides the natural framework for thermal quantum metrology.

A key feature of this representation is that $H$ is diagonal in the Bell basis. Consequently, the Gibbs state remains Bell diagonal for all values of the inverse temperature, allowing the concurrence, thermal quantum Fisher information, and curvature of entanglement to be obtained exactly in closed form. This also highlights an important distinction between coherent and equilibrium parameter encoding: in the present thermal setting the Bell eigenbasis remains fixed, while the parameter dependence enters entirely through the thermal populations.

The eigenstates of $H$ are the Bell states
\begin{equation}
\ket{\beta_{ab}}
=
\frac{1}{\sqrt2}
\left(
\ket{0,b}
+
(-1)^a
\ket{1,1\oplus b}
\right),
\qquad a,b\in\{0,1\},
\end{equation}
with corresponding eigenvalues
\begin{equation}
\omega_{ab}
=
(-1)^a \eta_x
-
(-1)^{a+b}\eta_y
+
(-1)^b\eta_z .
\label{eq:omega}
\end{equation}

We consider the Gibbs state at inverse temperature $\beta$,
\begin{equation}
\rho(\beta)
=
\frac{e^{-\beta H}}{Z(\beta)}
=
\sum_{a,b}
p_{ab}(\beta)\,
\ket{\beta_{ab}}\bra{\beta_{ab}},
\label{eq:gibbs}
\end{equation}
with Boltzmann weights
\begin{equation}
p_{ab}(\beta)
=
\frac{e^{-\beta\omega_{ab}}}{Z(\beta)},
\qquad
Z(\beta)
=
\sum_{a,b}
e^{-\beta\omega_{ab}}.
\label{eq:Z}
\end{equation}

The partition function may be written in the compact form
\begin{equation}
Z(\beta)
=
2\left[
e^{-\beta\eta_z}
\cosh\!\bigl(\beta(\eta_x-\eta_y)\bigr)
+
e^{\beta\eta_z}
\cosh\!\bigl(\beta(\eta_x+\eta_y)\bigr)
\right].
\label{eq:Zcompact}
\end{equation}

Physically, lowering the temperature enhances the population imbalance among the Bell levels, driving the system toward the Bell-state ground state. As shown below, this thermal population structure governs both the onset of thermal entanglement and the relation between entanglement curvature, robustness, and metrological sensitivity.

\section{Thermal entanglement and threshold}
\label{sec:thermal_entanglement}

For two-qubit mixed states, concurrence provides an exact measure of bipartite
entanglement~\cite{Wootters1998}. In the special case of Bell-diagonal states,
the expression simplifies considerably: entanglement is determined entirely by
the largest Bell-state population~\cite{Horodecki1996BellDiagonal,Berry2006TwoQubitTemperature}. Since the
thermal state in Eq.~(\ref{eq:gibbs}) is Bell diagonal, its concurrence is
\begin{equation}
C_{\rm th}(\beta)=\max\!\left\{0,\,2p_{\max}(\beta)-1\right\},
\label{eq:Cth_def}
\end{equation}
where $p_{\max}(\beta)=\max_{a,b}p_{ab}(\beta)$.

This form has a simple physical interpretation. A Bell-diagonal state is an
incoherent mixture of maximally entangled Bell states. Such a mixture is
entangled only when one Bell component dominates strongly enough to overcome the
classical mixing with the remaining three components. Thus thermal entanglement
is not produced by coherence between energy eigenstates, but by population
imbalance among them.

For $\eta_x\ge \eta_y\ge \eta_z>0$, Eq.~(\ref{eq:omega}) shows that the lowest
eigenvalue of $H$ is
\begin{equation}
\omega_{11}=-\eta_x-\eta_y-\eta_z .
\end{equation}
Hence the dominant thermal population is the Bell-ground-state occupation,
\begin{equation}
p_{\max}(\beta)=p_{11}(\beta)
=
\frac{e^{-\beta \omega_{11}}}{Z(\beta)}
=
\frac{e^{\beta(\eta_x+\eta_y+\eta_z)}}{Z(\beta)} .
\label{eq:pmax}
\end{equation}
In the entangled branch, the concurrence therefore reduces to
\begin{equation}
C_{\rm th}(\beta)=2p_{11}(\beta)-1 .
\label{eq:Cth_branch}
\end{equation}

Thermal entanglement appears precisely when the Bell-ground-state population
exceeds one half,
\begin{equation}
p_{11}(\beta)>\frac{1}{2}.
\label{eq:threshold_simple}
\end{equation}
This threshold reflects the competition between interaction-induced ordering and
thermal mixing. Lowering the temperature (increasing $\beta$) enhances the
occupation of the lowest Bell level. Conversely, at high temperature the Bell
populations approach $1/4$, and the state becomes separable.

Using Eq.~(\ref{eq:pmax}) together with Eq.~(\ref{eq:Zcompact}), the threshold
condition $p_{11}=1/2$ becomes
\begin{equation}
e^{-\beta\eta_z}\cosh\!\bigl(\beta(\eta_x-\eta_y)\bigr)
+
e^{\beta\eta_z}\cosh\!\bigl(\beta(\eta_x+\eta_y)\bigr)
=
e^{\beta(\eta_x+\eta_y+\eta_z)} .
\label{eq:threshold_explicit}
\end{equation}
Equation~(\ref{eq:threshold_explicit}) defines the sharp boundary separating
separable and entangled thermal states. This boundary is the equilibrium
analogue of an entanglement onset condition: below it, thermal mixing washes out
Bell-state entanglement; above it, the dominant Bell population is sufficiently
large to produce nonzero concurrence. Similar thermal-entanglement thresholds
have been studied in two-qubit and spin-chain models with anisotropic exchange
interactions~\cite{Arnesen2001,Wang2001,Rigolin2004,Sun2003,Berry2006TwoQubitTemperature}. The analysis in the remainder of this work is restricted to the entangled branch,
$p_{11}(\beta)>\frac12,$ for which the concurrence is differentiable and given by Eq.~(\ref{eq:Cth_branch}). At the threshold $p_{11}=1/2$, the concurrence changes from zero to nonzero values and is therefore nonanalytic due to the maximization in Eq.~(\ref{eq:Cth_def}). Consequently, the curvature-based quantities introduced below are understood away from the threshold.

\section{Thermal quantum Fisher information}
\label{sec:thermal_qfi}

The quantum Fisher information (QFI) determines the ultimate precision bound for parameter estimation through the quantum Cramér--Rao inequality~\cite{BraunsteinCaves1994,Paris2009,TothApellaniz2014}. For a mixed state $\rho(\beta)$ depending on the inverse temperature $\beta$, the QFI is
\begin{equation}
F_Q(\beta)
=
2\sum_{i,j}
\frac{\left|\langle i|\partial_\beta\rho|j\rangle\right|^2}
{\lambda_i+\lambda_j},
\label{eq:qfi_general}
\end{equation}
where $\{\lambda_i,|i\rangle\}$ are the instantaneous eigenpairs of $\rho(\beta)$.

In coherent metrology, the parameter dependence typically enters through the evolution of the eigenvectors, leading to interference contributions to the QFI. In the present equilibrium setting, however, the Bell basis diagonalizes $\rho(\beta)$ for all values of $\beta$, so the eigenvectors remain fixed and the QFI reduces entirely to the classical Fisher information of the thermal populations,
\begin{equation}
F_Q^{\rm th}(\beta)
=
\sum_{a,b}
\frac{\bigl(\partial_\beta p_{ab}\bigr)^2}
{p_{ab}}.
\label{eq:qfi_classical}
\end{equation}

Using Eq.~(\ref{eq:Z}),
\begin{equation}
p_{ab}(\beta)
=
\frac{e^{-\beta\omega_{ab}}}
{Z(\beta)},
\end{equation}
one obtains
\begin{equation}
\partial_\beta p_{ab}
=
-\bigl(\omega_{ab}-\langle\omega\rangle\bigr)p_{ab},
\qquad
\langle\omega\rangle
=
\sum_{a,b}
p_{ab}\omega_{ab}.
\label{eq:dpdbeta}
\end{equation}

Substituting into Eq.~(\ref{eq:qfi_classical}) gives
\begin{equation}
F_Q^{\rm th}(\beta)
=
\sum_{a,b}
p_{ab}
\bigl(\omega_{ab}-\langle\omega\rangle\bigr)^2
=
\Var_{\rho(\beta)}(\omega),
\label{eq:FQth_final}
\end{equation}
where
\begin{equation}
\Var_{\rho(\beta)}(\omega)
=
\langle\omega^2\rangle
-
\langle\omega\rangle^2 .
\end{equation}

Equation~(\ref{eq:FQth_final}) shows that the thermal QFI is governed entirely by equilibrium fluctuations of the interaction Hamiltonian. Unlike coherent metrology, where sensitivity arises from phase accumulation and basis evolution, the present equilibrium setting is purely fluctuation driven: the distinguishability between nearby thermal states is determined by how strongly the Gibbs populations respond to changes in inverse temperature.

\section{Symmetric logarithmic derivative and optimal measurements}
\label{sec:SLD}

We now examine the structure of the symmetric logarithmic derivative (SLD) for the thermal family
\begin{equation}
\rho(\beta)=\frac{e^{-\beta H}}{Z(\beta)},
\end{equation}
where $H$ is the two-qubit interaction Hamiltonian introduced in Sec.~\ref{sec:model}. Because the Gibbs state is a function of $H$, the density matrix commutes with the Hamiltonian for all values of $\beta$,
\begin{equation}
[\rho(\beta),H]=0 .
\end{equation}

Differentiating the Gibbs state gives
\begin{equation}
\partial_\beta\rho
=
-\bigl(H-\langle H\rangle\bigr)\rho ,
\end{equation}
where
\begin{equation}
\langle H\rangle=\mathrm{Tr}[\rho H].
\end{equation}
Comparing this expression with the defining SLD equation
\begin{equation}
\partial_\beta\rho
=
\frac12\bigl(L_\beta\rho+\rho L_\beta\bigr),
\end{equation}
one immediately obtains
\begin{equation}
L_\beta
=
-\bigl(H-\langle H\rangle\bigr).
\label{eq:thermalSLD}
\end{equation}

Since $H$ is diagonal in the Bell basis, the SLD shares the same eigenvectors. Consequently, the eigenbasis of $L_\beta$ is the Bell basis for all values of $\beta$. This has a direct operational implication: the projective measurement that saturates the quantum Cramér--Rao bound corresponds to a Bell-basis measurement, which is an entangled measurement on the two qubits.

Importantly, the optimal measurement basis is independent of the parameter $\beta$. This behavior contrasts with the coherent encoding scenario studied in our earlier work~\cite{Saleem2025CoE}, where the optimal measurement basis can vary with the parameter and becomes separable at special points where the curvature of entanglement equals the quantum Fisher information. In the equilibrium setting considered here, the SLD eigenbasis remains fixed and entangled for all values of $\beta$. Thus the relation between curvature of entanglement and quantum Fisher information does not correspond to a simplification of the measurement strategy but instead reflects a property of the thermal spectral structure of the system.

\section{Thermal Speed of Entanglement}
\label{sec:thermal_speed}

Before examining curvature, it is useful to characterize the first-order response of thermal entanglement to changes in inverse temperature. We therefore define the thermal speed of entanglement as
\begin{equation}
S_{\rm th}(\beta)
\equiv
\frac{dC_{\rm th}}{d\beta}.
\label{eq:thermal_speed_def}
\end{equation}

For Bell-diagonal thermal states in the entangled branch, the concurrence is given by Eq.~(\ref{eq:Cth_branch}). Using Eq.~(\ref{eq:dpdbeta}), we obtain
\begin{equation}
S_{\rm th}(\beta)
=
2\frac{dp_{11}}{d\beta}
=
2p_{11}(\beta)
\left(
\langle\omega\rangle-\omega_{11}
\right).
\label{eq:thermal_speed}
\end{equation}

Since $\omega_{11}$ is the ground-state energy, $\langle\omega\rangle\ge\omega_{11}$ and therefore $S_{\rm th}(\beta)\ge0$. Thus thermal entanglement increases monotonically as the temperature is lowered. The rate of increase is determined by the occupation probability of the dominant Bell state and its separation from the thermal average energy.

The speed of entanglement is bounded by the thermal quantum Fisher information. Using Eq.~(\ref{eq:FQth_final}) and the inequality
$\left(\langle\omega\rangle-\omega_{11}\right)^2
\le
\Var_{\rho(\beta)}(\omega)$,
one obtains
\begin{equation}
\left|S_{\rm th}(\beta)\right|
\le
2p_{11}(\beta)\sqrt{F_Q^{\rm th}(\beta)}
\le
2\sqrt{F_Q^{\rm th}(\beta)}.
\label{eq:thermal_speed_bound}
\end{equation}

The thermal quantum Fisher information therefore places an upper bound on the rate at which thermal entanglement can change with inverse temperature. This result is the thermal analogue of the information-geometric entanglement-speed bound derived for coherently evolving two-qubit states in Ref.~\cite{saleem2026quantumfisherinformationspeed}. Unlike the coherent setting, where entanglement speed is controlled by phase accumulation, the thermal speed is entirely population driven and reflects the redistribution of Gibbs weights among the Bell eigenstates.

Having characterized the linear response of thermal entanglement, we now turn to the curvature of entanglement, which quantifies the second-order response of concurrence to changes in inverse temperature.

\section{Thermal curvature of entanglement}
\label{sec:thermal_coe}

The curvature of entanglement (CoE) characterizes how the entanglement changes under variations of the estimated parameter~\cite{Saleem2025CoE}. In the present thermal setting, we define
\begin{equation}
\CoE(\beta)
\equiv
-\frac{d^2 C(\beta)}{d\beta^2}.
\label{eq:coe_def}
\end{equation}

For Bell-diagonal thermal states, the concurrence in the entangled branch is given by Eq.~(\ref{eq:Cth_branch}). Hence
\begin{equation}
\CoE^{\rm th}(\beta)
=
-2\frac{d^2 p_{11}}{d\beta^2}.
\label{eq:CoEth_start}
\end{equation}

Using Eq.~(\ref{eq:dpdbeta}),
\begin{equation}
\partial_\beta p_{ab}
=
-\bigl(\omega_{ab}-\langle\omega\rangle\bigr)p_{ab},
\end{equation}
together with
\begin{equation}
\frac{d\langle\omega\rangle}{d\beta}
=
-\Var_{\rho(\beta)}(\omega),
\label{eq:domegadbeta}
\end{equation}
one obtains
\begin{equation}
\frac{d^2 p_{11}}{d\beta^2}
=
p_{11}
\left[
\bigl(\omega_{11}-\langle\omega\rangle\bigr)^2
-
\Var_{\rho(\beta)}(\omega)
\right].
\end{equation}

The thermal curvature of entanglement therefore becomes
\begin{equation}
\CoE^{\rm th}(\beta)
=
2p_{11}(\beta)
\left[
\Var_{\rho(\beta)}(\omega)
-
\bigl(\omega_{11}-\langle\omega\rangle\bigr)^2
\right].
\label{eq:CoEth_final}
\end{equation}

Using Eq.~(\ref{eq:FQth_final}), this may be written as
\begin{equation}
\CoE^{\rm th}(\beta)
=
2p_{11}(\beta)\,F_Q^{\rm th}(\beta)
-
2p_{11}(\beta)
\bigl(\omega_{11}-\langle\omega\rangle\bigr)^2.
\label{eq:CoE_FQ_relation}
\end{equation}

Since the second term is nonnegative, we immediately obtain
\begin{equation}
\boxed{
\CoE^{\rm th}(\beta)
\le
2p_{11}(\beta)\,F_Q^{\rm th}(\beta).
}
\label{eq:state_bound}
\end{equation}

Because $0\le p_{11}(\beta)\le1$, this further implies
\begin{equation}
\boxed{
\CoE^{\rm th}(\beta)
\le
2F_Q^{\rm th}(\beta).
}
\label{eq:thermal_bound}
\end{equation}

It shows that, in thermal equilibrium, the curvature of entanglement is bounded by the same quantity that determines the precision of inverse-temperature estimation. Unlike the coherent case considered in Ref.~\cite{Saleem2025CoE}, where nontrivial saturation conditions arise from constructive interference between interaction-frequency sectors, the thermal bound is generally strict. Equality can occur only when the thermal average energy coincides with the dominant Bell-state energy, which for a nondegenerate spectrum implies vanishing thermal fluctuations and hence $F_Q^{\rm th}=\mathrm{CoE}^{\rm th}=0$.

Physically, Eq.~(\ref{eq:CoEth_final}) shows that the curvature of entanglement is determined by two competing quantities. The first is the variance of the interaction spectrum, which also determines the thermal quantum Fisher information. The second depends on the separation between the dominant Bell-state energy and the thermal average energy. The bound in Eq.~(\ref{eq:thermal_bound}) follows because this second contribution is always nonnegative.

The curvature of entanglement also admits a simple statistical interpretation. Writing the concurrence as
\begin{equation}
C_{\rm th}=2\langle\Pi_{11}\rangle-1,
\qquad
\Pi_{11}=|\beta_{11}\rangle\langle\beta_{11}|,
\end{equation}
and using the Gibbs-state relation
\begin{equation}
\partial_\beta\rho
=
-(H-\langle H\rangle)\rho,
\end{equation}
one finds
\begin{equation}
\CoE^{\rm th}(\beta)
=
-2\,\Cov_\rho\!\bigl(\Pi_{11},(\delta H)^2\bigr),
\qquad
\delta H=H-\langle H\rangle .
\end{equation}
Thus the curvature of entanglement is determined by connected correlations between the occupation of the dominant Bell state and energy fluctuations.

\section{Robustness of thermal entanglement generation}

Thermal entanglement arises from the imbalance of Gibbs populations among the Bell eigenstates. As the inverse temperature increases, the population of the Bell-state ground level grows and the thermal concurrence increases. Any uncertainty in the inverse temperature modifies these populations and therefore changes the amount of thermal entanglement present in the system. It is therefore natural to ask how sensitive thermal entanglement is to imperfect knowledge or control of the temperature.

To address this question, we consider fluctuations of the form
\begin{equation}
\beta \rightarrow \beta+\xi ,
\end{equation}
where $\xi$ is a random variable satisfying
\begin{equation}
\langle \xi \rangle =0,
\qquad
\langle \xi^2 \rangle = \sigma_\beta^2 .
\end{equation}
The experimentally observed concurrence is then
\begin{equation}
\overline{C}_{\rm th}(\beta)
=
\int d\xi\, p(\xi)\,
C_{\rm th}(\beta+\xi).
\end{equation}

Expanding the concurrence about $\beta$ and averaging over the fluctuations yields
\begin{equation}
\overline{C}_{\rm th}(\beta)
=
C_{\rm th}(\beta)
+
\frac{\sigma_\beta^2}{2}
\frac{d^2 C_{\rm th}}{d\beta^2}
+
O(\sigma_\beta^3).
\end{equation}

Motivated by the reduction in concurrence caused by temperature uncertainty, we define the thermal robustness
\begin{equation}
R_\beta(\beta)
=
\frac{
C_{\rm th}(\beta)
-
\overline{C}_{\rm th}(\beta)
}
{\sigma_\beta^2}.
\label{eq:Rbeta}
\end{equation}
Using the definition of the curvature of entanglement, Eq.~(\ref{eq:coe_def}), we obtain
\begin{equation}
\lim_{\sigma_\beta\rightarrow 0}
R_\beta(\beta)
=
\frac{1}{2}
\mathrm{CoE}_{\rm th}(\beta).
\label{eq:RbetaCoE}
\end{equation}

Equation~(\ref{eq:RbetaCoE}) shows that the curvature of entanglement determines the leading-order reduction of thermal entanglement caused by uncertainty in the inverse temperature. Regions where the concurrence varies rapidly with temperature are therefore the most susceptible to temperature fluctuations.

Combining Eq.~(\ref{eq:RbetaCoE}) with the bound in Eq.~(\ref{eq:thermal_bound}) immediately gives
\begin{equation}
\boxed{
\lim_{\sigma_\beta\rightarrow 0}
R_\beta(\beta)
\le
F_Q^{\rm th}(\beta).
}
\label{eq:robustnessbound}
\end{equation}
Equivalently,
\begin{equation}
C_{\rm th}(\beta)
-
\overline{C}_{\rm th}(\beta)
\le
\sigma_\beta^2
F_Q^{\rm th}(\beta),
\end{equation}
to leading order in the temperature uncertainty.

Equation~(\ref{eq:robustnessbound}) provides a direct connection between thermal entanglement and thermal metrology. The thermal quantum Fisher information determines the ultimate precision with which the inverse temperature may be estimated, while at the same time bounding the leading-order loss of thermal entanglement caused by uncertainty in that temperature. The bound shows that thermometric sensitivity limits the robustness of thermal entanglement.

This behavior differs from the coherent setting considered in Ref.~\cite{saleem2026information}. There, uncertainty modifies the phases accumulated during unitary evolution and affects entanglement through interference between different interaction-frequency sectors. Under special conditions, the resulting robustness bound can be saturated, so that the quantum Fisher information directly quantifies the loss of entanglement. In thermal equilibrium no such dynamical phases exist. Instead, temperature uncertainty acts by modifying the Gibbs populations that determine the thermal concurrence. Because the thermal quantum Fisher information depends on the response of all thermal populations, whereas the concurrence depends only on the dominant Bell-state population, the thermal robustness bound is generally not saturable. Consequently, part of the thermometric sensitivity of the Gibbs state is not visible through thermal entanglement.

From a practical perspective, Eq.~(\ref{eq:robustnessbound}) should therefore be viewed as a conservative estimate of entanglement loss. Given an independently measured or calculated thermal quantum Fisher information and an estimate of the inverse-temperature variance $\sigma_\beta^2$, one immediately obtains an upper bound on the expected reduction in concurrence. In general, the actual loss of thermal entanglement will be smaller because a portion of the thermal sensitivity is carried by population changes that do not contribute to the concurrence.

\section{Discussion}
We have investigated the relationship between thermal entanglement and quantum metrology in interacting two-qubit Gibbs states. For the most general anisotropic interaction admitting a Bell-diagonal representation, we obtained exact expressions for the thermal concurrence, thermal quantum Fisher information, and the first and second derivatives of concurrence with respect to inverse temperature. These results establish a direct connection between thermal entanglement and equilibrium fluctuations.

The analysis reveals important differences from the coherent setting studied in Refs.~\cite{saleem2026quantumfisherinformationspeed,Saleem2025CoE}. While coherent dynamics can exhibit nontrivial saturation of information-geometric bounds through interference effects, the corresponding thermal bounds are generally strict. The reason is that the quantum Fisher information depends on the response of the entire Gibbs distribution, whereas the concurrence depends only on the occupation of the dominant Bell state. Consequently, not all thermometric information contained in the thermal state is visible through entanglement alone.

We further showed that the speed and curvature of thermal entanglement are constrained by the thermal quantum Fisher information, and that temperature uncertainty leads to a reduction in thermal entanglement bounded by the same quantity. These results extend the connection between entanglement response and quantum metrology from coherent dynamics to equilibrium quantum systems and provide a unified perspective on thermal entanglement, thermometric sensitivity, and robustness to temperature uncertainty. An interesting direction for future work is to investigate whether similar information-geometric constraints govern entanglement response in many-body thermal systems, particularly near quantum and thermal critical points where the quantum Fisher information can become strongly enhanced \cite{Zanardi2007,Gu2010}. It would also be interesting to explore possible applications to quantum thermometry and criticality-enhanced sensing, where equilibrium fluctuations play a central role \cite{Correa2015OptimalThermometry,DePasquale2016ThermalSusceptibility,Mehboudi2019}.

\bibliographystyle{apsrev4-2}
\bibliography{bibliography}

\end{document}